\documentclass[aps,prl,twocolumn,showpacs,superscriptaddress]{revtex4}

\usepackage{graphicx} % Include figure files

\begin{document}

\title
{Transition Metal-Ethylene Complexes as High-Capacity
Hydrogen Storage Media
}

\author{E. Durgun}
\affiliation{Department of Physics, Bilkent University, Ankara
06800, Turkey} \affiliation{UNAM - National Nanotechnology
Research Center, Bilkent University, Ankara 06800, Turkey}
\author{S. Ciraci}
\email{ciraci@fen.bilkent.edu.tr} \affiliation{Department of
Physics, Bilkent University, Ankara 06800, Turkey}
\affiliation{UNAM - National Nanotechnology Research Center,
Bilkent University, Ankara 06800, Turkey}
\author{W. Zhou}
\affiliation{NIST Center for Neutron Research, National Institute
of Standards and Technology, Gaithersburg MD 20899}
\affiliation{Department of Materials Science and
Engineering, Univ. of Pennsylvania, Philadelphia, PA 19104}
\author{T. Yildirim}
\affiliation{NIST Center for Neutron Research, National Institute
of Standards and Technology, Gaithersburg MD 20899}
\affiliation{Department of Materials Science and
Engineering, Univ. of Pennsylvania, Philadelphia, PA 19104}

\date{\today}

\begin{abstract}
From first-principles calculations, we predict  that a single
ethylene molecule can form a stable complex with two transition
metals (TM) such as Ti. The resulting TM-ethylene complex then
absorbs up to ten hydrogen molecules, reaching to gravimetric
storage capacity of  $\sim$14 wt\%. Dimerization, polymerizations
and incorporation of the TM-ethylene complexes in nanoporous
carbon materials have been also discussed. Our results are quite
remarkable and open a  new approach to high-capacity hydrogen
storage materials discovery.

\end{abstract}

\pacs{73.63.Nm, 72.25.-b, 75.75.+a}

\maketitle

Hydrogen is considered as one of the best alternative and
renewable fuels\cite{cell,dresselhaus} due its abundance, easy
synthesis, and non-polluting nature when used in fuel cells.
However, the main concern is the safe storage and efficient
transport of this highly flammable gas\cite{review3}.

%Many types of materials have been tried or suggested
%for use as hydrogen-storage media.
%These include nano-materials, high surface area materials,
%synthetic metals, chemical and metal hydrides and chlathrates.
%Unfortunately none of these state-of-the-art materials currently
%satisfy the guidelines issued to make a commercially
%useful hydrogen storage system for competitive
%hydrogen-powered transportation.
%Therefore hydrogen storage in new materials and
%devices is an active field of research worldwide.

The main obstacles in hydrogen storage  are slow kinetics, poor
reversibility and high dehydrogenation temperatures for the
chemical hydrides\cite{bogdanovic}; and very low desorption
temperatures/energies for the physisorption materials
(metal-organic frameworks (MOF)\cite{mof_prl}, carbide-derived
carbons (CDC)\cite{cdc_jacs}, etc). Recently,  a novel concept to
overcome these obstacles have been
suggested\cite{taner1,taner2,heben,dag,heben2,nurten,jena18e,polymer}.
It was predicted that a single Ti-atom affixed to carbon
nanostructures, like C$_{60}$ or nanotubes, strongly adsorbs up to
four hydrogen molecules\cite{taner1,taner2,dag}. The interaction
between hydrogen molecules and transition metals is very unique,
lying between chemi and physisorption, with a binding energy of
0.4 eV compatible with room temperature desorption/absorption.
%At large Ti coverage,  it is possible to store hydrogen
%molecules up to 8-wt%.
The origin of this unusual "molecular chemisorption" is explained
by well-known Dewar coordination and Kubas
interaction\cite{kubas}. The transition metals (TM) are chemically
bonded onto different molecules/nanostructures through
hybridization of lowest-unoccupied molecular orbital (LUMO) of
nanostructure  with TM $d$-orbitals (i.e.  Dewar coordination).
The resulting complex then binds  multiple molecular hydrogens
through hybridization between H$_2-\sigma^*$ antibonding and TM
$d$-orbitals (i.e. Kubas interactions).

\begin{figure}
\includegraphics[width=6cm]{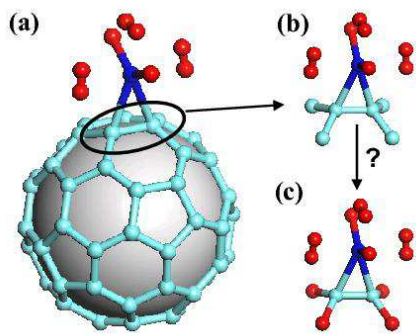}
\caption{(color online) (a) One of the most stable
structure of Ti-C$_{60}$ complex  where Ti atom (blue) is bonded
to a double bond with four  hydrogen molecules attached (red).
(b) The  local structure of  the Ti-C$_{60}$-double bond.
(c) Replacing the end carbon atoms
 shown in (b) by H results in an ethylene molecule. This suggests that we
 may simple use ethylene molecule to hold Ti atoms, which then  binds
 multiple hydrogen molecules.
 }
\label{fig:figure1}
\end{figure}

Synthesizing the predicted structures of Ti decorated
nanotubes/C$_{60}$ were proven to be very difficult due to lack of
bulk quantities of small-diameter nanotubes and strong
C$_{60}$-C$_{60}$ interactions in the solid phase. Moreover, Ti
atoms uniformly coating the SWNT/C$_{60}$ surface may be subject
to clustering after several charging-discharging
process\cite{jena}.

In  search for a more efficient and feasible high
capacity hydrogen storage medium,
we found that the C=C double bond of an ethylene molecule
C$_{2}$H$_{4}$, mimics double bonds of C$_{60}$  that strongly binds
TM-atom (see Fig. 1) and therefore it is
expected to support TM atoms strongly to provide a basis for
high-capacity hydrogen storage via the Dewar-Kubas mechanism discussed
above.

In this letter, we explored this idea and indeed found that a
single ethylene molecule can hold not only one but two  Ti atoms,
i.e. C$_{2}$H$_{4}$Ti$_2$, which then reversibly binds up to ten
H$_{2}$ molecules yielding an unexpectedly high storage capacity
of $\sim$ 14 wt \%. These results suggest that ethylene, a
well-known inexpensive molecule, can be an important basis in
developing frameworks for efficient and safe hydrogen storage
media.

Our results are obtained from first-principles plane wave
calculations within density functional
theory
%\cite{kohn}
using Vanderbilt-type ultra-soft pseudopotentials with
Perdew-Zunger exchange correlation\cite{pwscf}. Single molecules
have been treated in a supercell of 15$\times$15$\times$15 \AA~
with $\Gamma$ k-point and a cutoff energy of 408 eV.
%The supercell size of 15 \AA~ is large enough to prevent interactions
%between adjacent cells.
The structures are optimized
until the maximum force allowed on each atom is  less than
0.01 eV/$\AA$ for both spin-paired and spin-relaxed
cases.

We first studied  the bonding of a single Ti atom to an ethylene
molecule to form C$_2$H$_4$Ti (see Fig.~2(b)). We found no energy
barrier for this reaction. The binding energy is calculated  by
subtracting the equilibrium total energy $E_{T}$ of
C$_{2}$H$_{4}$Ti molecule from the sum of the total energies of
free ethylene molecule and of Ti atom;
E$_{B}$(Ti)=E$_{T}$(C$_{2}$H$_{4}$)+E$_{T}$(Ti)-E$_{T}$(C$_{2}$H$_{4}$Ti).
The Ti atom forms a symmetric bridge bond with the C=C bond of
ethylene with $E_{B}=$1.45 eV. Interestingly, it is also possible
to attach a second Ti atom to the C$_2$H$_4$Ti to form
C$_2$H$_4$Ti$_2$ (see Fig.~2(c)) without any potential barrier and
about the same binding energy as the first Ti atom. In the
optimized structure (Fig.~2(c)), each Ti atom is  closer to one of
the carbon atoms, leading to two different Ti-C bonds. Fig.~2(d)
shows that the bonding orbital for the Ti-atoms and C$_2$H$_4$
results from the hybridization of the LUMO orbital of the ethylene
molecule and the Ti-$d$ orbitals, in accord with Dewar
coordination\cite{kubas}.  The spin-polarized calculation gives
1.53 eV lower energy than non-spin-polarized one, suggesting
magnetic ground state for C$_{2}$H$_{4}$Ti$_{2}$ with  moment
$\mu=6 \mu_{B}$ per molecule.

\begin{figure}
\includegraphics[width=7cm]{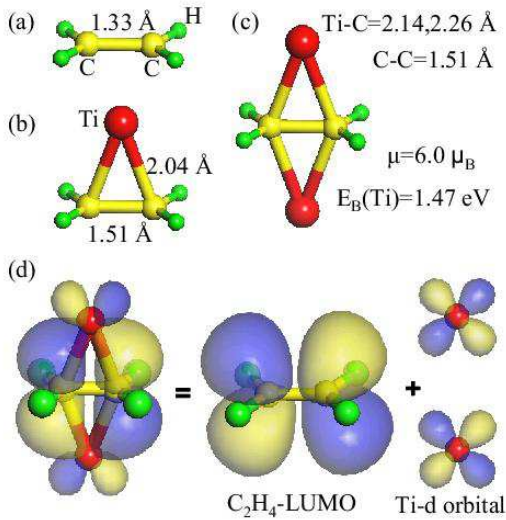}
\caption{
(Color online) Optimized structure of an ethylene
molecule C$_2$H$_4$ (a),  C$_2$H$_4$Ti (b), and
C$_2$H$_4$Ti$_2$ (c). The panel (d) shows that
Ti-C$_2$H$_4$ bonding orbital results from the
hybridization of the
LUMO of the C$_2$H$_4$ and the  Ti-$d$ orbital,
in accord with Dewar coordination.}
\label{fig:titoethy}
\end{figure}

The stability of the C$_{2}$H$_{4}$Ti$_{2}$ complex were further
tested by normal mode analysis. We found no soft (i.e negative)
mode. There are three main Ti-modes. In two of these  modes, Ti
atoms vibrate parallel and perpendicular to the C=C bond and have
the energies of 176 cm$^{-1}$ and 123 cm$^{-1}$, respectively. In
the third Ti-mode Ti atoms vibrate perpendicular to the C$_2$H$_4$
plane with energy of 367 cm$^{-1}$. These three modes are unique
for the C$_2$H$_4$Ti$_2$ complex and therefore should be present
in any Raman/IR spectra of a successfully synthesized material.
%%% TTTYYYY too much details for Ti2-c2h4--need more space
%Finally, we checked that
%C$_{2}$H$_{4}$Ti$_{2}$ molecule is also energetically more
%favorable by 0.34 eV against the desorption of Ti$_{2}$ dimer
%leaving behind a free ethylene molecule.

We next studied the H$_{2}$ storage capacity of Ti-ethylene
complex, by calculating the interaction between
C$_{2}$H$_{4}$Ti$_{2}$ and different number of H$_2$ molecules and
configurations. The first H$_2$ molecule is absorbed
dissociatively to form C$_{2}$H$_{4}$(TiH$_2$)$_2$ as shown in
Fig.~3(a) with a binding energy of 1.18 eV/H$_2$. The additional
hydrogen molecules do not dissociate and molecularly absorbed
around the Ti-atom. Two of these configurations are shown in
Fig.~3(b) and (c). In C$_{2}$H$_{4}$(TiH$_2$-2H$_2$)$_2$
configuration, two H$_2$ are molecularly bonded from left and
right side of the TiH$_2$ group with a binding energy of 0.38
eV/H$_2$ and significantly elongated   H-H bond length of 0.81
\AA. It is also energetically favorable to add a third H$_2$
molecule from the top of the  TiH$_2$ group, with a binding energy
of 0.4 eV and bond length of 0.82 \AA. The resulting structure,
C$_{2}$H$_{4}$(TiH$_2$-3H$_2$)$_2$, is shown in Fig.~3(c). We note
that these binding energies have the right order  of magnitude for
room temperature storage. Since the hydrogens are absorbed
molecularly, we also expect fast absorption/desorption kinetics.

Finally, we  also observed many other local stable configurations
where all of the hydrogen molecules are bonded molecularly. One of
such configuration, denoted as C$_{2}$H$_{4}$(Ti-5H$_2$)$_2$, is
shown in Fig.~3(d). Here the H$_2$ molecules stay intact and
benefit equally from bonding with the Ti atom. The total ten
hydrogen molecules absorbed by a single Ti-ethylene complex,
C$_{2}$H$_{4}$(Ti-5H$_2$)$_2$, corresponds to a $\sim$ 14 wt \%
gravimetric density, which is more than twice of the criterion set
for efficient hydrogen storage. By removing the top H$_2$
molecule, we find that C$_{2}$H$_{4}$(Ti-4H$_2$)$_2$ configuration
is also a local minimum and have a slightly (0.08 eV/H$_2$) higher
energy than the C$_{2}$H$_{4}$(TiH$_2$-3 H$_2$)$_2$ configuration
shown in Fig.~3(c). Our MD simulations indicate that the system
oscillates between these two configurations. This is consistent
with the observation that in Kubas compounds\cite{kubas}, the
dihydrogen (i.e 2H) and molecular  (i.e H$_2$ ) bonding are
usually found to be in resonance\cite{kubas}.

The top hydrogen molecule in
C$_{2}$H$_{4}$(Ti-5H$_2$)$_2$, has the weakest bonding
in the system with E$_B$=0.29 eV while the
side H$_2$ molecules have the strongest bonding with
E$_B$=0.49 eV/H$_2$ and significantly elongated
H-H bond distance of 0.85 \AA.  This suggests
the presence of two different
 H$_2$-C$_{2}$H$_{4}$Ti$_2$ bonding
orbitals as shown in  Fig~3(e) and (f). The first one is the
hybridization between the top-H$_2$  $\sigma^*$-antibonding and
the Ti-$d$ orbital. The second one is the simultaneous
hybridization of the side H$_2$ $\sigma^*$-antibonding orbitals
with the Ti-$d$ orbital. Since the bonding orbitals are  mainly
between metal $d-$  and hydrogen $\sigma^*$-antibonding orbitals,
the mechanism of this interesting interaction  can be explained by
the Kubas interaction\cite{kubas}.

\begin{table*}[htbp]
\caption{The binding energies (in eV)
with respect to atomic
and bulk energies of  various metals (M). The last two rows
indicate the maximum number of H$_2$ molecule bonded to each metal
and its average binding energy (in eV). }
\begin{center}
\begin{tabular}{c|ccccccccccccccc} \hline
Property/M & Sc & Ti & V & Cr & Mn &
Fe & Co & Ni & Cu & Zn & Zr & Mo & W & Pd & Pt \\ \hline
E$_B$ (M-atomic) & 1.39 & 1.47 & 1.27 & 0.05  & 0.37 & 0.83 &
1.30 & 0.70 & 1.41 & none & 1.69 & 0.37 &
1.18 & 1.56 & 1.78 \\
E$_B$(M-bulk) & -2.72 & -3.66 & -4.13 & -3.57 & -3.20 &
-1.74 &-2.53 & -2.19 & -2.25 & -- &-4.44 &-5.84 & -7.18 &-2.24 &-3.56 \\
max H$_2$/M
& 5
& 5
& 5
& 5
& 5
& 5
& 3
& 2
& 2
& --
& 5
& 5
& 5
& 2
& 2   \\
E$_B$(per H$_2$) & 0.39 &  0.45 &  0.43 & 0.35 &  0.34 &  0.26 &
0.41 & 0.87 & 0.14 & -- &  0.57 &  0.77 &  0.90 &  0.58
&  0.95 \\
\hline
\end{tabular}
\label{tab1}
\end{center}
\end{table*}

\begin{figure}
\includegraphics[width=7cm]{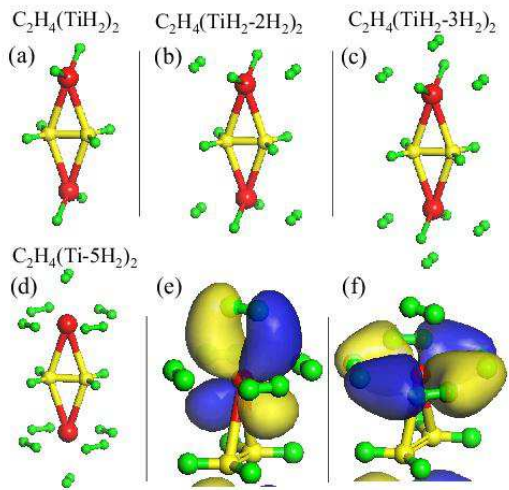}
\caption{
(Color online) Atomic configurations of an
ethylene molecule functionalized by two Ti atoms,
holding (a) two H$_2$ molecules which are dissociated,
(b) six H$_2$ molecules,
and (c) eight H$_2$ molecules.
Panel (d) shows a configuration where ten H$_2$
are bonded all molecularly. The spin-polarized
calculations gave lower energies (in eV) by 1.5,
0.37,0.16, and 0.06 for configurations shown in
(a)-(d), respectively, suggesting magnetic ground
state for all cases with moment of $\mu \approx 2 \mu_B$.
The panel (e) and (f) shows the bonding orbital for the top (e) and
side hydrogen molecules, respectively. Note that the
hydrogen $\sigma^{*}$-antibonding orbitals are hybridized
with Ti-$d$-orbitals, suggesting Kubas interaction for the
H$_2$-Ti bonding.}
\label{fig:etyhlene}
\end{figure}

We also calculated the normal modes of
C$_{2}$H$_{4}$(Ti-5H$_2$)$_2$ and did not find any soft modes,
indicating that the system indeed corresponds to a local minimum.
Among many vibrational modes, we note that the H$_2$ stretching
mode is around 2700-3000 cm$^{-1}$ for side H$_2$ and around 3300
cm$^{-1}$ for top H$_2$ molecules, significantly lower than the
4400 cm$^{-1}$ for free H$_2$ molecule. Such a shift in the mode
frequency would be the key feature that can be probed by Raman/IR
measurement to confirm a successful synthesis of the structures
predicted here.

Finally, the stability of C$_{2}$H$_{4}$(Ti-5H$_2$)$_2$
structure has
been further studied by carrying out \textit{ab-initio} molecular
dynamic calculations performed at 300~K and 800~K. While all ten
H$_{2}$ molecules have remained bound to C$_{2}$H$_{4}$Ti$_2$ molecule at
300~K, they start to desorb above 300~K and all of them are desorbed
already at 800~K, leaving behind a stable C$_{2}$H$_{4}$Ti$_{2}$
molecule. This prediction suggests that all the stored hydrogen
molecules can be discharged easily through heating.

\begin{figure}
\includegraphics[scale=0.4]{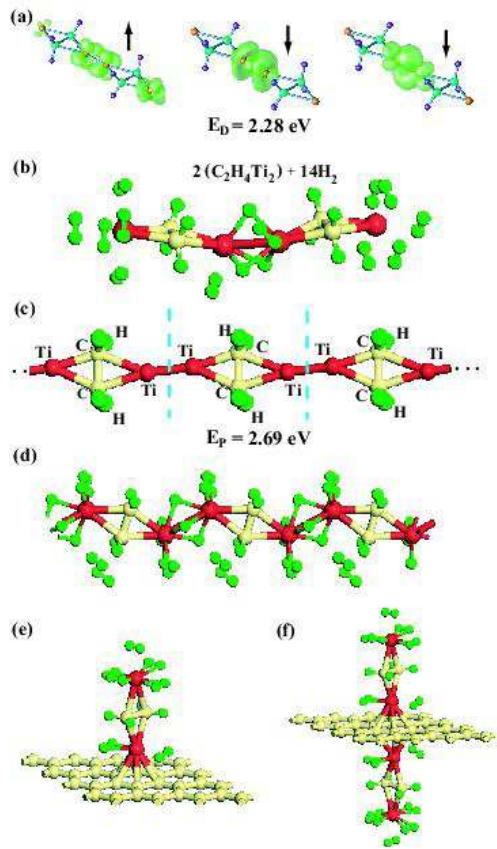}
\caption{(color online) (a) Atomic configuration and charge
density plots of a dimer of C$_{2}$H$_{4}$Ti$_{2}$ molecules
linked by two Ti atoms one from either molecule. Large, medium and
small balls indicate  titanium, carbon and hydrogen atoms,
respectively. Charge density plots (from left to right) correspond
to HOMO spin-up, and two spin-down states. (b) Atomic
configuration of C$_{2}$H$_{4}$Ti$_{2}$ dimer holding 14 H$_{2}$
molecules,
 (c) Polymer of C$_{2}$H$_{4}$Ti$_{2}$
molecule with polymerization energy $E_{P}$. (d) Atomic
configuration of the polymer holding 6 H$_{2}$ molecule. (e) A
C$_{2}$H$_{4}$Ti$_{2}$ molecule adsorbed above the center of a
hexagon in a (3$\times$3) cell holding 7 H$_{2}$. (f) Two
Ti-ethylene complexes adsorbed above and below the center of the
same hexagon holding 14 H$_{2}$.} \label{fig:twoethy}
\end{figure}

We next discuss the possibility of  the dimerization and
polymerization in the course of recycling and its effect on the
hydrogen storage capacity.  We found that two molecules can form a
dimer through a Ti-Ti bond as shown in Fig. \ref{fig:twoethy}(a).
The dimer formation is exothermic with an energy gain of
$E_{D}$=2.28 eV, and leads to a stable structure. The ground state
is ferromagnetic with $\mu=$6$\mu_{B}$. While each Ti atoms at
both ends of the dimer can bind 5 H$_{2}$ molecules, two linking
Ti atoms can absorb only 4 H$_{2}$
%(two of them dissociated and the remaining two molecularly absorbed)
totaling to 14 H$_{2}$ per dimer. As a result, the gravimetric
density obtained by the dimer is lowered to $\sim$ 10 wt\%. The
total energy can be further lowered by adding more
C$_{2}$H$_{4}$Ti$_{2}$ molecules to the dimer, and eventually by
forming a paramagnetic polymer as shown in Fig.
\ref{fig:twoethy}(c). The polymerization energy per molecule is
calculated to be $E_{P}$=2.69 eV  and the H$_{2}$ storage capacity
is further lowered to 6.1 wt\%. Polymerization did not change the
binding energy of the H$_2$ on Ti-atom significantly and therefore
should not affect the desorption temperature.

In order to prevent C$_{2}$H$_{4}$Ti$_{2}$ molecules to form a
possible polymer phase during recycling,  one can imagine to
incorporate these TM-ethylene complexes in a nanoporous materials
such as MOF\cite{mof_prl}
 and carbide-derived carbons\cite{cdc_jacs}.
As an example, we consider a C$_{2}$H$_{4}$Ti$_{2}$ adsorbed above
and below the center of a hexagon of a graphene layer. Here the
graphene is taken as a prototype system which represents the
internal structure of a carbon-based nanoporous material. We find
that TM-ethylene complex can form stable structure with graphene
surface. As shown in Fig. \ref{fig:twoethy} (e) and (f), single
and double C$_{2}$H$_{4}$Ti$_{2}$ molecules assembled on a
3$\times$3 graphene layer can hold 7 and 14 H$_{2}$ with average
binding energy of 0.43 and 0.41 eV/H$_2$, respectively. The
binding energy of the C$_{2}$H$_{4}$Ti$_{2}$ molecule is found to
be $\sim$2 eV. The actual binding energy in nanoporous materials
could be even higher due to curvature effects\cite{gulseren}.  The
maximum gravimetric density achieved in this present framework is
6.1 wt\%

It is important to know if the results reported above for M=Ti
hold for other metals. Therefore we are currently  studying  a
large number of metals and the details will be published
elsewhere\cite{wzhou}. Our initial results are summarized in
Table~I, which clearly indicates that most of the light TM atoms
can be bound to ethylene and each of them can absorb up to 5 H$_2$
molecules. Scandium is the ideal case but for practical reasons Ti
is the the best choice of elements. Cr binds very weakly while Zn
does not bind at all to the C$_2$H$_4$ molecule. Interestingly, Zr
forms stronger bonding with C$_2$H$_4$ than Ti and can absorb up
to ten H$_2$ molecularly with an average binding energy of 0.6 eV.
Heavier metals such as Pd and Pt can also form complexes with
C$_2$H$_4$ but bind less hydrogen molecules with significantly
stronger binding energy than Ti. Table~I  also gives the binding
energies with respect to bulk metal energies. The negative value
for E$_B$ indicates endothermic reaction. Due to very low vapor
pressure of many metals, it is probably better to use some
metal-precursor  to synthesize the structures predicted here.

In conclusion, we showed that an individual ethylene molecule
functionalized by two light transition metals can bind up to ten
hydrogen molecules via Dewar-Kubas interaction, reaching  a
gravimetric density as high as $\sim$14 wt\%. We propose to
incorporate TM-ethylene complex into carbon-based nanoporous
materials to avoid dimerization/polymerization during recycling.
Our results open a  new approach to high-capacity hydrogen storage
materials discovery by  functionalizing small organic molecules
with light transition metals.

%Apart
%from high H$_{2}$ storage capacity, ethylene gains interesting
%magnetic properties through functionalization by Ti and retain
%very high energy when it is dressed by H$_{2}$.

\begin{acknowledgments}
This work was partially supported by T\"{U}B\.{I}TAK under Grant
No. TBAG-104T536. W.Z. and  T.Y. acknowledge partial support from
DOE under DE-FC36-04GO14282  and BES grant DE-FG02-98ER45701. We
thank Dr Sefa Dag for fruitful discussions.
\end{acknowledgments}

\end{document}